\documentclass[square]{ws-procs11x85}
\usepackage{multicol,float} 

\begin{document}

\title{Parsec-scale Investigation of the Magnetic Field Structure of Several AGN Jets}

\author{S. P. O'Sullivan and D. C. Gabuzda}

\address{Department of Physics, University College Cork, Ireland\\
E-mail: shaneosullivan@physics.ucc.ie\\
gabuzda@physics.ucc.ie}

\begin{abstract}
Multi-frequency (4.6, 5, 5.5, 8, 8.8, 13, 15, 22 \& 43 GHz) polarization observations of 6 ``blazars'' were obtained on the American Very Long Baseline Array (VLBA) over a 24-hr period on 2 July 2006.
Observing at several frequencies, separated by short and long intervals, enabled reliable determination of the distribution of Faraday Rotation on a range of scales.
In all cases the magnitude of the RM increases in the higher frequency observations, implying that the electron density and/or magnetic field strength is increasing as we get closer to the central engine.
After correcting for Faraday rotation, the polarization orientation in the jet is either parallel or perpendicular to the jet direction.
A transverse Rotation Measure (RM) gradient was detected in the jet of 0954+658, providing evidence for the presence of a helical magnetic field surrounding the jet.
For three of the sources (0954+658, 1418+546, 2200+420), the sign of the RM in the core region changes in different frequency-intervals, indicating that the line-of-sight component of the magnetic field is changing with distance from the base of the jet. We suggest an explanation for this in terms of bends in a relativistic jet surrounded by a helical magnetic field; where there is no clear evidence for pc-scale bends, the same effect can be explained by an accelerating/decelerating jet.
\end{abstract}


\bodymatter

\begin{multicols}{2}

\section{Introduction}
AGN jets emit synchrotron radiation that can be detected at all radio
frequencies. This radiation is often highly linearly polarized, which
provides important information on the degree of order and
orientation of the magnetic field in these jets.
The type of AGN studied in this experiment are known as ``blazars'',
which have jets pointed close to our line of sight (LoS) and often
exhibit strong variability in total flux and linear polarization over a
broad range of frequencies from $\gamma$-ray to radio.

We observed 6 sources with the American Very Long Baseline Array (VLBA) at 8 frequencies from
4.6 GHz to 43 GHz over a 24-hr period on 2 July 2006. This radio interferometer provides milliarcsecond
resolution which corresponds to the parsec scale structure of these
jets. Even though these AGN are intrinsically two-sided, pc-scale
observations typically show a one sided ``core-jet'' structure due to
Doppler boosting of the radiation from the relativistic jet
pointed towards us. The jet moving away from us is usually too faint
to be detected.

The radiation detected from the jet is generally optically thin
while the core region usually displays a flat spectrum attributed to
synchrotron self-absorption. In the optically thin regime the
polarization orientation is perpendicular to the magnetic field, while in
the optically thick/flat spectrum region the polarization is parallel to the
magnetic field.

Recent MHD simulations have provided an almost complete explanation of how these jets are launched, accelerated and collimated close to the black hole, see \cite{MeierJapan} and references therein. The
global magnetic field structure expected on these scales is helical. It is possible that this
structure changes when the flow gets disrupted by shocks after a few
hundred Schwarzschild radii, but observational evidence of helical
fields on scales larger than this \cite{Asada2002, GabuzdaMurray2004, Mahmud2008} suggests that remnants of the
earlier magnetic field structure remain or that a current driven helical kink
instability is generated \cite{NakamuraJapan, Carey2008}.

\section{Faraday Rotation}
A rotation of the plane of polarization of an electromagnetic wave
occurs when it propagates through a region with free electrons and
a magnetic field, commonly referred to as Faraday Rotation. The effect
is manifest as a linear dependence of the polarization angle
($\chi$) on the wavelength ($\lambda$) squared and is described by the formula
\begin{equation}
\chi_{obs}=\chi_0+\left(\frac{e^{3}}{2\pi m^2c^4}\int{n_e}{\mathbf{\vec{B}.d\vec{l}}}\right)\lambda^2
\end{equation}
where the Rotation Measure (RM), in parentheses, is proportional to the integral of
the electron density and the dot product of the magnetic field along
the LoS with the path length $\vec{dl}$. Hence, a positive/negative RM
tells you that the LoS magnetic field is pointing
towards/away from the observer.

For a pure $e^{+}/e^{-}$ plasma one would expect no
Faraday rotation because of cancellation due to the $e^3$ dependence.
The dependence on the inverse square of the mass also shows that the
effect is much greater for thermal electrons than for protons or
relativistic electrons.
Therefore, multi-frequency observations enable us to see gas that
would otherwise be invisible to us and, combined with the corrected
polarization orientation, we are provided with a 3-D view of the
magnetic field structure.

\begin{figure}[H]
\centerline{\psfig{file=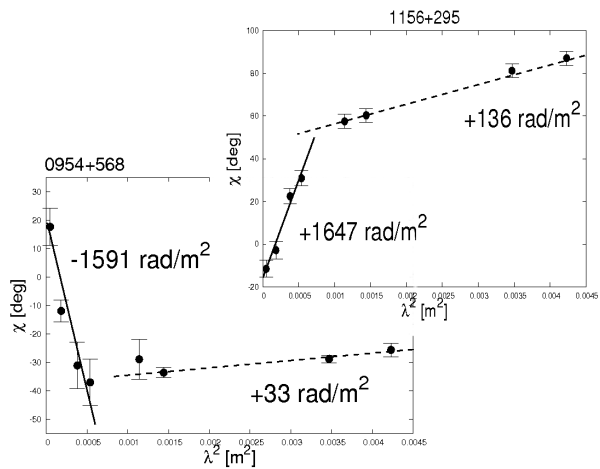,width=8cm}}
\caption{Plots of $\chi$ vs. $\lambda^2$ for the whole observed frequency range of the core regions of 1156+295 (right) and 0954+658 (left) clearly showing a transition in both cases where physically different regions are being probed}\label{aba:fig1}
\end{figure}

However, if the rotating medium is mixed in with the emitting plasma then
internal Faraday rotation occurs and the above formula does not generally apply.
Theoretical models \cite{Burn1966, Cioffi1980} of internal FR in a spherical or cylindrical
region have shown that, if a rotation of greater than 45 degrees is
observed, 
the Faraday rotation must be external. Many observations show rotations greater than 45 degrees,
\cite{GabuzdaMurray2004, ZTFog2, Gab2006} and our current observations, indicating that the rotating gas must be segregated
from the emitting region. Furthermore, observations of short
term RM variability \cite{ZTvar2001} means that the Faraday rotating gas must
be very close to the source, most likely in a boundary layer or sheath surrounding the jet.

\section{Observations}
The calibration and imaging was performed with the radio astronomy software
package AIPS. The integrated foreground (mainly Galactic) RMs were subtracted
using the multi-frequency VLA observations of \cite{Pushk2001}. This isolates the
RM distribution in the immediate vicinity of the AGN and is very important
for source rest frame calculations, since the observed RM scales with $(1+z)^2$.

Our current observations are ideally suited for Faraday rotation analysis,
with both long and short spaced frequency intervals, which significantly
reduces the $\pm n\pi$ ambiguity in the observed polarization angles. This enables
us to construct RM maps on a range of different scales and Faraday depths
providing a wealth of information on the RM distribution and sign in the core
region and out along the jet. Matched-resolution images were constructed for
different frequency intervals where clear transitions in RM occurred, corresponding to sampling
of physically different regions (eg. Fig. 1). 

\begin{figure}[H]
\centerline{\psfig{file=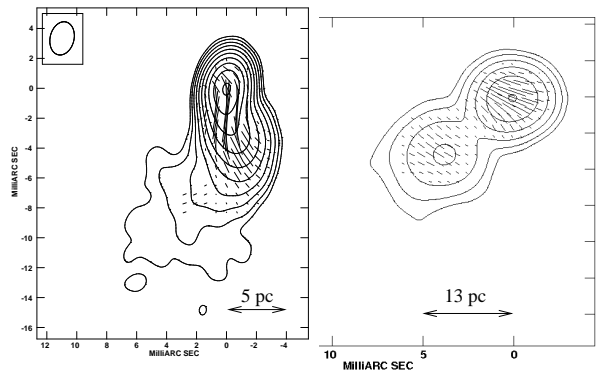,width=8cm}}
\caption{The left image of 2200+420 clearly shows how the polarization sticks are aligned with the jet direction and, importantly, remain aligned even as the jet bends. The image on the right of 1418+546 is the one source that displays polarization perpendicular to the jet direction. The contours correspond to total radio flux while the sticks correspond to the polarization orientation.}\label{aba:fig2}
\end{figure}

\section{Results}
Since Faraday rotation has a very significant effect on the observed
polarization angle, $\chi_{obs}$, the rotation must be removed in order to obtain the
intrinsic polarization orientation, $\chi_0$, and reach any definite conclusions about the
magnetic field structure. From our multi-frequency observations
we find that the polarization in the optically thin jet is either aligned with
(0954+658, 1156+295, 2007+777, 2200+420) or perpendicular to (1418+546)
the observed jet direction (eg. Fig. 2).
\footnote{No appreciable jet polarization was detected in 1749+096}
The aligned polarization cases can be attributed to the toroidal component of
a relatively tightly wound helical field, whereas the poloidal component, which
dominates in a loosely wound helical field, can account for the one
perpendicular polarization observation \cite{Lyutikov2005}.

\begin{table}[H]
\tbl{Note the higher magnitude RM in the high $\nu$ range for all sources, as well as the RM \emph{sign} change in half the sources. All RM values in $rad/m^2$}
{\begin{tabular}{@{}ccccr@{}}\toprule

Blazar & z & Core RM & Core RM \\
 & & (Low $\nu$ range)& (High $\nu$ range) \\ \colrule

0954+658 & $0.368$ & $+33\pm14$ & $-1591\pm265$\\
1156+295 & $0.729$ & $+136\pm4$ & $+1647\pm159$\\
1418+546 & $0.152$ & $+75\pm8$ & $-501\pm48$\\
1749+096 & $0.320$ & $-33\pm24$ & $-500\pm347$\\
2007+777 & $0.342$ & $+638\pm39$ & $+1946\pm140$\\
2200+420 & $0.069$ & $+240\pm90$ & $-1008\pm43$\\ \botrule

\end{tabular}}\label{aba:tbl1}
\end{table}

In all cases, the RM has a larger magnitude in the core region than in the jet
for a particular frequency interval. The magnitude of the core RM also increases in
the higher frequency intervals for all sources, simply implying that the
electron density and/or magnetic field strength is increasing as we sample regions
closer to the central engine. This can be clearly seen in Table 1 where a summary
of the core RM results is presented. The redshift, z, is also included.

\begin{figure}[H]
\centerline{\psfig{file=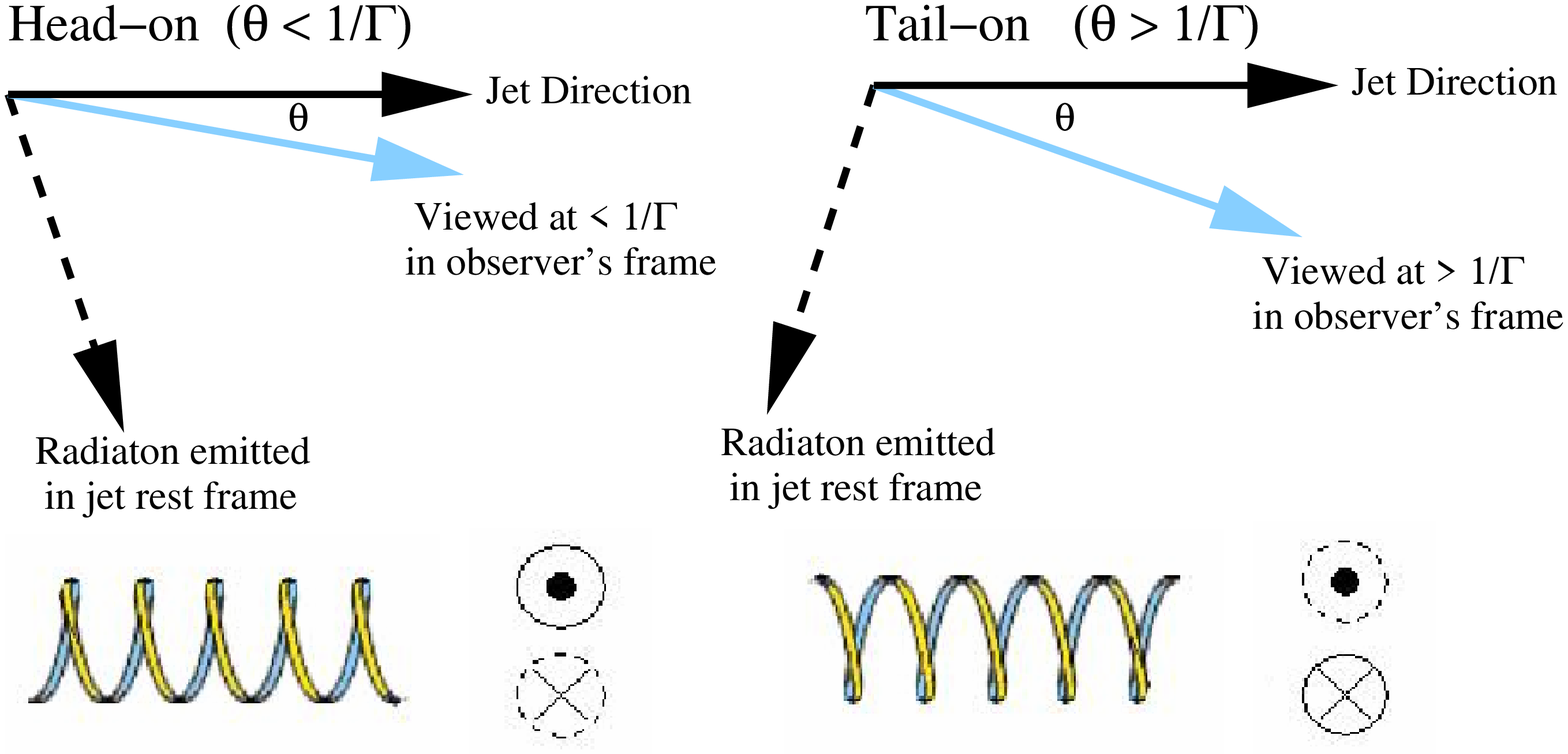,width=8cm}}
\caption{For a ``head-on'' ($\theta<1/\Gamma$) or ``tail-on'' ($\theta>1/\Gamma$) view of a helical magnetic field, different LoS components will dominate (shown by solid circles surrounding the dot or cross which indicate the direction of the LoS component) producing RMs with different signs in an unresolved jet.}\label{aba:fig3}
\end{figure}

A surprising result is that, for three of the sources, the \emph{sign} of the RM
changes in different frequency intervals, telling us that the LoS
component of the magnetic field is changing with distance from the base of
the jet.
In the case of 2200+420, the dominant jet magnetic field is transverse
throughout the jet, even as it bends (see Fig. 2), consistent with the presence of a global
helical magnetic field structure. Bends in a jet, surrounded by a helical magnetic
field, on scales smaller than is probed by the lowest frequency resolution will produce different dominant LoS magnetic field components due to the relativistic motion of the jet towards us, see Fig. 3. This effect provides a natural explanation of the observed RM sign changes.

\begin{figure}[H]
\centerline{\psfig{file=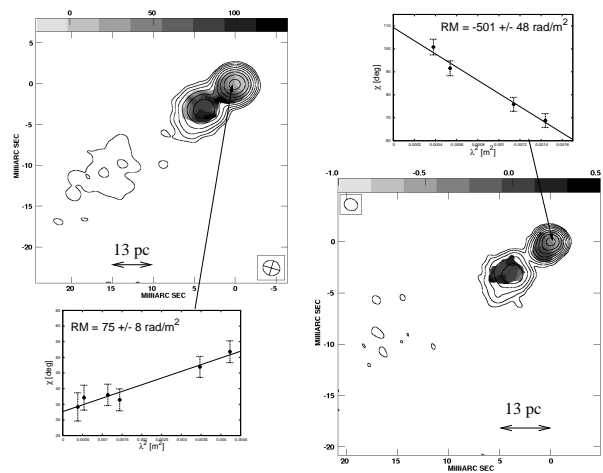,width=8cm}}
\caption{RM maps of 1418+546 in different frequency intervals showing how the sign of the RM changes in the core. Grey scale denotes the RM in $rad/m^2$ and contours are of total radio flux. The insets are plots of $\chi$ vs. $\lambda^2$ for the corresponding regions.}\label{aba:fig4}
\end{figure}


If there is no clear evidence for bends in the inner jet, as is the case for 0954+658 and
1418+546 (eg. Fig. 4), then the same effect can occur for an accelerating/decelerating jet
surrounded by a helical magnetic field; although it remains possible that bends are present on
scales smaller than our resolution.

Further analysis of 0954+658 reveals the presence of a RM gradient across the
jet (Fig. 5), which is a strong signature for the presence of a helical magnetic field
geometry. For a side-on view of a helical magnetic field, one would expect a
systematic gradient across the jet with the LoS magnetic field changing sign.
However, as was shown in \cite{Asada2002}, a same-sign gradient (as
detected in 0954+658) can be explained by having the viewing angle ($\theta$) in the jet
rest frame smaller than the pitch angle ($\psi$) of the helix.

Another excellent diagnostic of the magnetic field structure are transverse
profiles of the jet. For a helical magnetic field with $\theta<\psi$, one would expect the peak
in the polarization profile to be offset from the peak of the total flux
profile, and furthermore, the magnitude of the RM gradient should increase
on the side opposite to the total flux peak offset \cite{Papa2008}. Transverse profiles of 
the jet of 0954+658 display this pattern, see Fig. 5.

\begin{figure}[H]
\centerline{\psfig{file=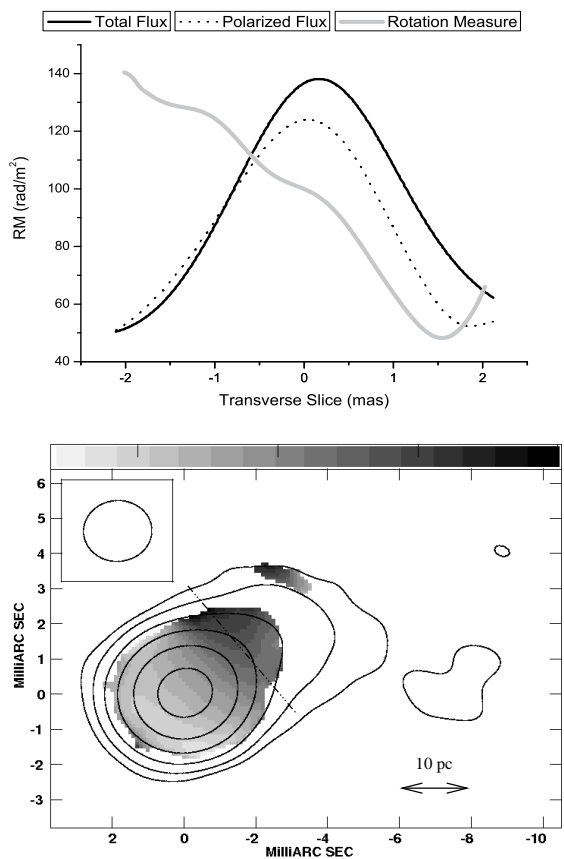,width=8cm}}
\caption{The grey scale image displays the RM gradient in 0954+658 with accompanying scaled transverse profiles of total flux (solid line), polarized flux (dotted line) and RM intensity (grey line) across the slice indicated.}\label{aba:fig5}
\end{figure}

\section{Conclusions}
Several phenomenon, for example, the polarization structure \cite{Lyutikov2005}, RM gradients \cite{Asada2002} and circular polarization generation \cite{GabCP2007}, can be understood by considering a helical magnetic field geometry for pc-scale AGN jets.
In this paper, we consider how regions with different RM signs can also be explained within a helical magnetic field model, as places where the jet is observed at angles greater than or less
than $1/\Gamma$ due to bends in the jet or due to an accelerating/decelerating straight jet. (A longitudinal jet magnetic field with a change in the angle to the LoS could also cause a RM sign change, but this does not correspond to the observed magnetic field in 0954+658 or 2200+420.)
 It's important to note that VLBI resolution is usually not
sufficient to completely resolve the true optically thick core, therefore, the VLBI ``core'' consists
of emission from the true core and some of the optically thin inner jet. So if bends occur on
scales smaller than the observed VLBI ``core'', core RMs with different signs can be derived
from observations at different frequencies (ie. probing different scales of the inner jet).

In our future work, we will attempt to reconstruct the 3-D path of the jet through space
using the combined information from the observed distributions of the total intensity, linear polarization, spectral index and rotation measure.

\end{multicols}
\end{document}